# Signals vs. Videos: Advancing Motion Intention Recognition for Human-Robot Collaboration in Construction

## Charan Gajjala Chenchu[1], Kinam Kim[2*], Gao Lu[2], and Zia Ud Din[2]

[1]Department of Computer Science, University of Houston, Texas, USA
[2]Department of Construction Management, University of Houston, Texas, USA
*Corresponding Author

cgajjala@cougarnet.uh.edu, kkim48@central.uh.edu, lgao5@central.uh.edu, uziauddi@central.uh.edu

**Abstract -**

Human-robot collaboration (HRC) in the construction industry depends on precise and prompt recognition of human motion intentions and actions by robots to maximize safety and workflow efficiency. There is a research gap in comparing data modalities—signals and videos—for motion intention recognition. To address this, the study leverages deep learning to assess two different modalities in recognizing workers' motion intention at the early stage of movement in drywall installation tasks. The Convolutional Neural Network–Long Short-Term Memory (CNN-LSTM) model utilizing surface electromyography (sEMG) data achieved an accuracy of around 87% with an average time of 0.04 seconds to perform prediction on a sample input. Meanwhile, the pre-trained Video Swin Transformer combined with the transfer learning technique, harnessed video sequences as input to recognize motion intention and attained an accuracy of 94% but with a longer average time of 0.15 seconds for a similar prediction. This study emphasizes the unique strengths and trade-offs of both data formats, directing their systematic deployments to enhance HRC in real-world construction projects.

**Keywords -**

Human-Robot Collaboration (HRC); Motion Intention Recognition; Surface Electromyography (sEMG); Convolutional Neural Network–Long Short-Term Memory (CNN-LSTM); Video Swin Transformer

## 1 Introduction

Human-robot collaboration (HRC) refers to the mutual coordination between humans and robots in a collaborative workspace, where robots can understand the environmental context, and human actions and can make decisions to supplement human tasks. Due to recent advancements in hardware, software, and deep learning algorithms, the performance of construction robots has increased in construction projects in terms of safety, efficiency, and productivity of construction workers [1]. Despite the advancements, construction robots still face challenges due to the dynamic and rapidly changing environments where understanding human behaviors with precise timing is essential [2]. Hence, it is critical to recognize workers' motion intention to enable HRC in construction [3].

Human motion intention recognition is being researched and implemented by modeling machine learning algorithms on two primary modalities: sensor and video data. Among sensor-based approaches, surface electromyography (sEMG) sensors have been utilized to capture the muscle activity of human because they represent muscle movements and their acquisition is easier due to their non-intrusive nature [4]. Meanwhile, videos are also being extensively utilized to analyze human motion with emerging computer vision and image processing technologies [5]. Videos can be used to predict the motion trajectories of workers, equipment, and harmful actions to enable safety monitoring in construction sites [6, 7].

However, these two modalities have their limitations. EMG signals are prone to noise and variability across different subjects which make generalizing patterns difficult [4]. It may be uncomfortable for workers to carry sensor with them. On the other hand, videos require high computational power for model training and inference. Video-based approaches are vulnerable to occlusions, varying viewpoints, background clutters, etc. There have been studies exploring these approaches separately, but the existing studies lack a solid comparison of these modalities in this recognition task, especially in construction.

To address this research gap, this study proposes a comprehensive comparison of the applications of sEMG and video-based modalities as input to deep learning algorithms in human motion intention recognition in construction. This research implements a Convolutional Neural Network–Long Short-Term Memory (CNN-LSTM) model to analyze sEMG signals and utilizes a pre-trained Video Swin (Shifted Window) Transformer model to train on video input. Finally, the performance of these two models is evaluated and compared. This proposed system could provide insights into tradeoffs between their performance and drawbacks, which can help to choose an appropriate approach among the two options for HRC in construction.

## 2 Literature Review

To enable adaptive and seamless cooperation between human workers and robotic systems, the construction industry has centered on integrating human motion recognition and behavioral analysis into HRC. This included sensors, wearable devices, computer vision, and machine learning models, to capture and interpret human movements in real-time [1, 8]. This section reviews the state-of-the-art studies in human motion recognition using two primary modalities and their applications in construction.

### 2.1 Signal-based Methods

sEMG and other signal-based methods are extensively adopted in industrial and construction applications [3]. These methods can predict human movements before they occur which can enhance safety and operational efficiency of exoskeleton robots [9]. sEMG signals were also used to recognize hand gestures and control industrial robots in the context of a smart factory [10].

Shah and Kim [3] proposed an sEMG-based motion intention recognition system at an early stage by analyzing the first 1-second muscle activity from four muscle sensors. They implemented a CNN-LSTM model and achieved an accuracy of 98.5% in classifying four construction-related activities. Combining convolutional layers with gated recurrent unit layers (GRU), a hybrid model was trained on sEMG signals [10]. This model was tested on 11 subjects and obtained around 96% accuracy in detecting 10 types of hand gestures. A CNN-LSTM-based system was developed to predict various motion intentions in a squat activity which represents lifting motion in industrial environments [9]. This methodology can catch the movement initiations 0.3 seconds ahead using sEMG signals with an accuracy of 84.02%.

While sEMG plays a key role in human motion detection, several studies show that these signals are prone to variability created by factors such as electrode placement, muscle fatigue, skin impedance, environmental weather conditions, and subject-level bias in terms of gender and age, which leads to noise that can hamper accuracy in action detection [4, 11].

### 2.2 Video-based Methods

Tang et al. [7] presented a methodology for utilizing the LSTM model's encoder-decoder architecture combined with multi-head prediction modules to predict the future paths of workers and equipment on construction sites using video data. A transformer-based model [6] was proposed for recognizing unsafe gestures on construction sites which applied transformer and 3D CNN to extract spatial and temporal features. It achieved a precision of 88.7% on a customized dataset outperforming baseline methods. For swimming action recognition, a novel multi-modal approach called Swintrans Net [12] was presented, building upon the Swin-Transformer's [13] superior feature extraction and adaptability.

Video-based motion recognition systems on construction sites also encounter several challenges. Blocked camera views by frequent obstructions from materials, machinery, and workers, result in incomplete or inaccurate motion data. With constantly changing layouts, equipment, and the dynamic nature of these environments, it is difficult to develop robust systems capable of adapting to unpredictable conditions [2]. Real-time motion detection can be a technical problem due to the processing time of high-resolution videos. Additionally, varied lighting conditions, such as low light, shadows, and glare, can further degrade the performance [14].

### 2.3 Research Gap

Comparative analysis between sEMG and video data sources in motion intention recognition is relatively scarce, specifically in the context of HRC in construction. One potential reason is due to the limitation of publicly available datasets containing both sensor and video data types that represent similar experimental settings [5]. Two data types can be utilized in analyzing human motion in their own way, but they also have some shortcomings. This presents an opportunity to compare the two approaches in terms of dataset dimensions, deep learning models, accuracy, performance, and prediction latency. This systematic comparison can aid researchers in developing more effective motion intention recognition systems tailored to specific needs and tasks. For the sEMG, we chose a CNN-LSTM model since it outperforms non-recurrent models in time series analysis [15]. Video Swin Transformer model was chosen for its advantage over 3D CNNs in capturing long-range spatial and temporal dependencies in video sequences using self-attention mechanisms [6, 12].

## 3 Proposed Methodology

The proposed methodology starts with data acquisition of raw sEMG signals and videos which are then preprocessed as inputs to the models. The obtained training curves and test evaluation metrics are interpreted for a comparative analysis. The system workflow is depicted in Figure 1.

### 3.1 Signal-based Approach

#### 3.1.1 Data Acquisition and Preprocessing

The sEMG signal data was captured from 3 subjects performing four drywall installation activities multiple times. This data consists of four channels representing signals coming from four EMG electrodes, which were attached to the biceps brachii, triceps brachii, shoulder, and forearm muscles of the dominant hand, respectively. These sensor placements are effective in capturing the muscle activity of these actions [3]. A baseline reference node was attached to the subjects on top of the wrist of the same hand. This data was stored in a text file in a tabular format in which rows are time series data points with each point containing 4 columns corresponding to the 4 channels. Hence, one such text file represents the muscle signals captured during one subject doing one of the drywall activities one time.

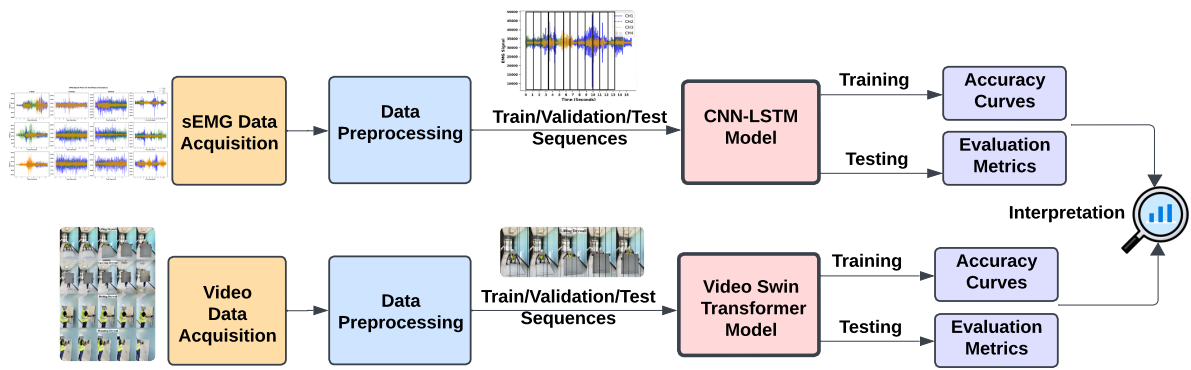

Figure 1. Overall System Architecture

The activities were divided into "Intention" group, which constitutes the first 1 second of the activity, and the rest as "Actual" group. The time series signals were divided into fixed-size sequences. A sequence represents a small fraction of continuous movements. Consequently, the motion detection task was translated into a classification task with 2 classes (corresponding to the two groups) per each activity, totaling 8 labels. In this way, the whole dataset of text files was divided into overlapping chunks with a sliding window approach. Due to the small length of the intention group, the number of chunks that belong to this intention group became significantly smaller compared to chunks in the actual group which can lead to a major class imbalance within the groups. To tackle this, minority intention classes were randomly oversampled by duplicating the existing samples by 2 times ensuring uniform distribution across activities. This dataset was then divided into train, validation, and test splits ensuring stratification for a uniform and representative distribution of each activity and subject type. A random Gaussian noise was added to the training set consistently to each data point in a sequence as part of the data augmentation. Consequently, these three sets were normalized using a standard scaler that transforms the data to have a mean of 0 and a standard deviation of 1.

### 3.1.2 Deep Learning Model

The proposed model architecture consists of two 1D CNN blocks followed by two LSTM layers and two fully connected dense layers at the end as shown in Figure 2. This hybrid model can process both spatial and temporal features inherent in sEMG data. The initial 1D CNN layers are designed for the extraction of local patterns in the form of spatial features from the multichannel signals. Next, the output activations are normalized across the batch using the batch normalization (Batch Norm) layer. After this, max-pooling layers are employed to retain only the most useful patterns. To mitigate overfitting, dropout layers are integrated to randomly nullify the contributions of a fraction of input units during training. This can make the model generalize well to the training data by preventing it from overly relying on specific hidden units. The second CNN block is added with an increase in filter size from 64 to 128 (see Figure 2). Following this second block, LSTM units are deployed to exploit their inherent feedback connections in the form of memory gates and cells [3]. This enables the analysis of temporal dependencies in the sEMG sequences without the problem of vanishing gradients. The multidimensional output from the LSTM layers is transformed into a 1D vector using a flatten layer before passing them to fully connected dense layers. Finally, these learned attributes are mapped to the 8 desired output classes using a dense layer with softmax activation.

### 3.2 Video-based Approach

### 3.2.1 Data Acquisition and Preprocessing

While the subjects performed the activities equipped with the sEMG sensors, videos were recorded using a typical smartphone at the same time. A single data point in this context refers to a single image frame. One video file corresponds to the action sequence generated by one subject performing one type of activity at a time. Each video was also partitioned into Intention and Actual activity groups with the first 1 second interval labeled as Intention classes. All the videos were resized to 224 × 244. The pixel values were normalized to have a range between 0 and 1 to improve model convergence and stability during training. These frames were further segmented into overlapping sequences that represent small fragments of continuous movements. The created chunks were divided into train, validation, and test splits such that they contained a representative distribution of sequences in terms of actions and subjects. The labeling strategy used in the signal-based approach was deployed in the video-based approach. Common image data augmentations were applied to the training dataset, ensuring all the frames in a sequence receive the same transformation. This sequence-level augmentation virtually increases training data size, so that models can generalize over varying noise, occlusions, and other real-world conditions.

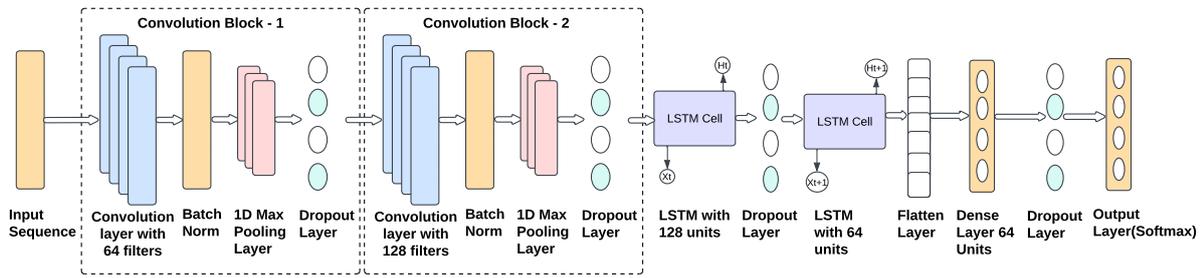

Figure 2. Proposed CNN-LSTM Model Architecture

#### 3.2.2 Deep Learning Model

In this study, the Video Swin Transformer architecture was implemented for the classification task from videos [16]. Initial tokenization followed by a patch merging layer maintains the hierarchical structure of the model. The Video Swin Transformer block [16] starts with a 3D window-based multi-head self-attention (MSA) module, which processes tokens within independent 3D windows to capture specific spatiotemporal relationships in the video. This block also employs a 3D shifted window approach between layers which integrates context from adjacent areas and frames. This lets it recognize long-running complex interactions and actions throughout the video input. To learn more detailed patterns, multilayer perceptron (MLP) is applied after the MSA module to transform features non-linearly.

The proposed approach uses the existing base version of the Video Swin model (Video Swin-B) which was pre-trained on a dataset called Kinetics 600 [16]. We employed a transfer learning strategy, starting with these pre-trained weights frozen for training. The final MSA module, layer normalization, and dense layers of the last transformer block in Stage 4 are unfrozen to update their weights (see Figure 3). However, the MLP layer was retained to be frozen to preserve the high-level non-linear feature transformations learned during pretraining. This makes the model fine-tune to only the attention-driven feature extraction process, reducing complexity and overfitting. Finally, a custom classification head is added that contains a dense output layer with 8 neurons (corresponding to the target classes) at the end, as shown in Figure 3. This head is added to scale complex representations extracted by the transformer and finally map the reduced features to class probabilities using softmax activation.

## 4 Experiment

### 4.1 Data Acquisition

We conducted an experiment that involved 3 subjects performing four drywall installation tasks as displayed in Figure 4. These tasks [3] were selected because of their potential to be supported by construction robots. Each subject performed these tasks four times with varying durations while carrying four sEMG muscle sensors and a reference electrode. Simultaneously, these activities were recorded using a smartphone camera. Using OpenSignals software, the sEMG signals (see Figure 4) were collected and stored in a ".txt" format at a frequency of 500 Hertz (Hz), implying that for one second 500 data rows, each with four channel columns would be produced. Videos were filmed in ".mp4" format at a resolution of 1920 × 1080 pixels with 60 frames per second (FPS). In this way, the original raw dataset of sensor signals contains 16 text files for each subject with 4 files per activity per subject, totaling 48 files for the 3 subjects. The video dataset also contains the same number of video files. Table 1 provides a comprehensive overview of the experimental setup including the raw data point counts.

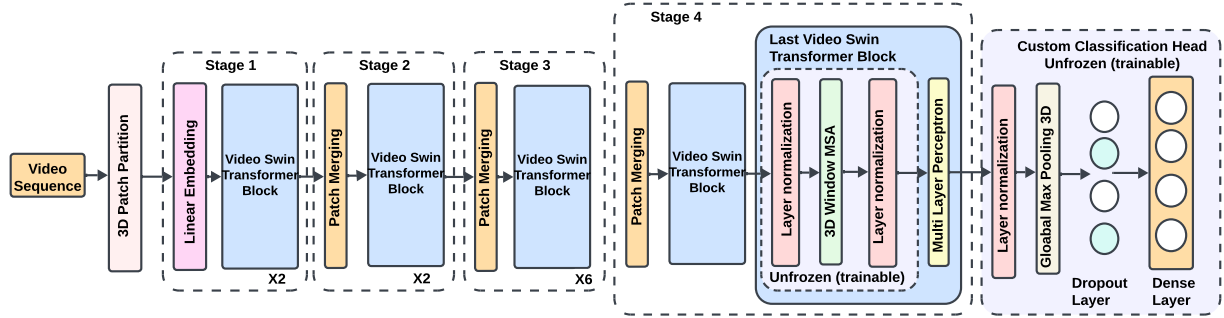

Figure 3. Video Swin Transformer (Base version) Architecture with Custom Classification Head

Table 1. Data Comparison Between Signal-Based and Video-Based Approaches

| Method | Data | File Format | Frequency | File Count | Data Point Count |
|---|---|---|---|---|---|
| Signal | sEMG signals | .txt | 500 Hz | 48 | 195,450 |
| Video | RGB Videos | .mp4 | 60 fps | 48 | 20,690 |

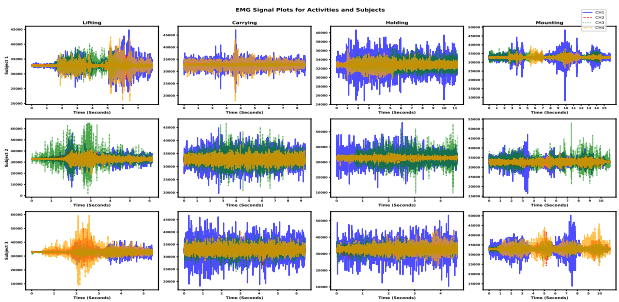 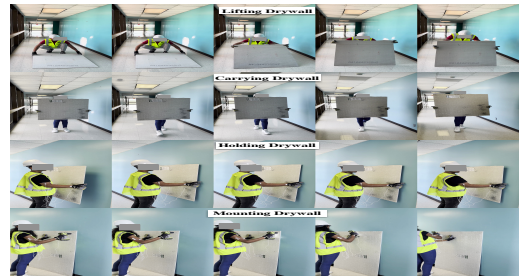

Figure 4. Sample sEMG Signals (Left) and Video Frames (Right) of activities performed by 3 subjects

### 4.2 Data Preprocessing

Following Shah and Kim [3], the first 1 second of each activity was regarded as the Intention group and the rest as the Actual group. As a result, 8 target classes, namely Lifting Intention, Actual Lifting, Carrying Intention, Actual Carrying, Holding Intention, Actual Holding, Mounting Intention, and Actual Mounting, were formed. This can enable robust recognition of motion triggers and their transition to actual motion. For implementation purposes, the intention groups from the video dataset were divided into separate video clips, producing a total of 96 videos. Due to the size of the sensor data points being significantly higher compared to videos, the video files were increased by 2 times using data augmentations, yielding a final dataset of 288 videos. The signal data was processed into sequences in each of 100 data rows, while each sequence extracted from videos contains 32 frames. Thus, a sequence of signals equates to 0.2 seconds of movement and a video sequence encloses 0.5 seconds of motion. Such an individual sequence becomes the single unit of input to the models. To increase the classification accuracy, both datasets were fragmented into overlapping sequences with a ratio of 50%. To overcome the imbalance between "Intention" and "Actual" classes, the intention sequences were oversampled 3 times for signal data, but the video sequences were kept unaltered since they were already increased.

The EMG sequences were split into train, validation and test splits in the ratio of 70:15:15 respectively. The individual video files were first divided into train, validation, and test splits in the ratio of around 67%, 17%, and 17% respectively, maintaining the representative distribution of activity and subject types. Random Gaussian noise was applied to signal training examples. For the video training dataset, along with Gaussian noise, image, augmentations like random horizontal flip, hue, brightness, saturation, and contrast were employed. Finally, the sensor data was scaled using a standard scaler whereas the image frames were resized to a resolution of 224 × 244 and then scaled by dividing pixel values by 255.

Table 2. Comparison of Data Preprocessing: Signal-Based vs. Video-Based Approaches

| Method | Intention Group | Resolution | Seq. Length | Overlap | Oversampling |
|---|---|---|---|---|---|
| Signal | 1 second | 1 row, 4 channels | 100 | 50% | 3x |
| Video | 1 second | 1 Frame (224x224, RGB) | 32 | 50% | None |

### 4.3 Training and Hyperparameter Tuning

The CNN-LSTM model was trained on the signals training dataset and the pre-trained Video Swin-B model was partially fine-tuned on the training video sequences. A dedicated validation dataset was used for guiding the model selection process due to its capability of offering an unbiased estimate of model performance. For both models, sparse categorical cross-entropy loss was used to handle the integer encoded true labels and Adam optimizer with fixed learning rate (LR) was employed to leverage the adaptive learning rate. Class weights were incorporated during training to tackle the imbalance across the activity classes. The majority classes were assigned lesser weights compared to the minority labels based on the inverse frequency of the classes and the loss was scaled dynamically using these computed weights. The sensor-based and video-based models were trained for 1000 and 100 epochs with batch sizes of 128 and 32 respectively. The intermediate dropout layers and the final dropout layer in the proposed CNN-LSTM model were given a dropout rate of 0.4. Ridge (L2) regularization was applied to the 1D convolutional and LSTM layers in the form of the kernel and recurrent regularizers, respectively. Analogously, an L2 rate of 0.1 and a dropout value of 0.5 were utilized for the last output and dropout layers in the custom head of the proposed video model. A grid search method was deployed to identify the best combinations of L2 rate and dropout values, with L2 ranging from 0.0001 to 0.1 in steps of 0.05 and dropouts from 0.1 to 0.5 in the same increments. The best hyper-parameter values are outlined in Table 3.

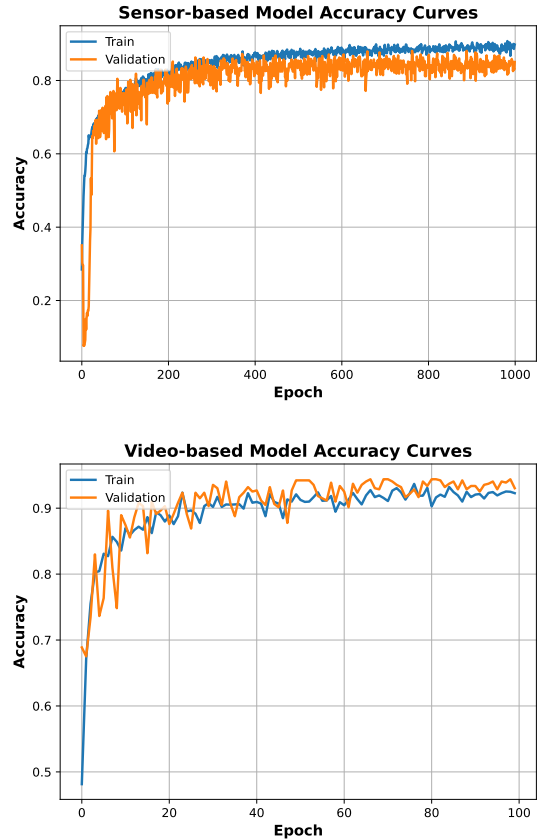

Figure 5. Training and Validation Accuracy curves for signal-based (top) and video-based models (bottom).

## 5 Results

The accuracies of training and validation data over the iterations were recorded and plotted as indicated in Figure 5. The video-based model achieved consistently higher accuracy, with training and validation curves stabilizing above 90%, indicating robust generalization and minimal performance fluctuations. On the other hand, the CNN-LSTM model exhibited a slower rise in accuracy, stabilizing around 85%, with higher variability in the validation accuracy curve. Overall, both the models were trained without overfitting.

After the training was completed, test sets were used to evaluate the models' performances. Based on test predictions, confusion matrices (Figure 6) were computed to assess the performance in individual classes. Due to the imbalance in the classes, overall weighted F1 scores were also determined. The Video Swin-B model demonstrates high accuracies for categories such as 'Actual Mounting' (95.72%), 'Actual Lifting' (98.1%), and 'Actual Holding' (100%). It also distinguished intentions effectively, correctly predicting all 18 instances of 'Carrying Intention' and 'Lifting Intention'. In contrast, the CNN-LSTM model shows greater variability in performance, with higher misclassifications. 'Actual Carrying' showed more confusion with 98 out of 129 instances being correct and 11 'Actual Lifting' instances being mislabeled as Lifting Intention. The model faced challenges in distinguishing intentions, as shown by the lower performances on 'Carrying Inten-tion' and 'Mounting Intention' classes.

Table 3. Training and Testing: Signal-Based vs. Video-Based Approaches

| Method | Model | LR | Epochs | Batch Size | Dropout | L2 | Accuracy | F1 Score |
|---|---|---|---|---|---|---|---|---|
| Signal | CNN-LSTM | 0.001 | 1000 | 128 | 0.4 | 0.05 | 86.82% | 86.79% |
| Video | Video Swin-B | 0.001 | 100 | 32 | 0.5 | 0.1 | 94.05% | 94.22% |

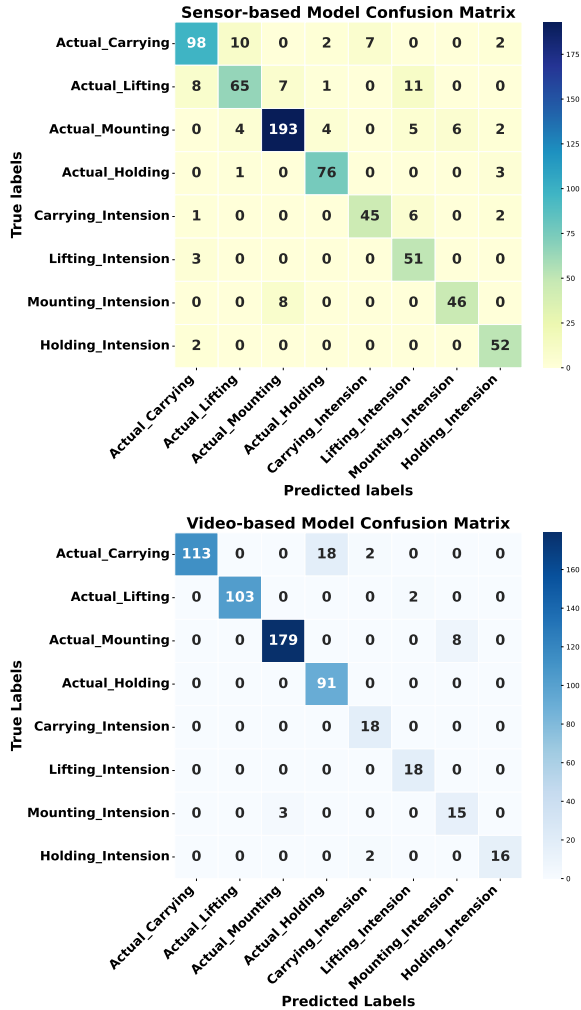

Figure 6. Confusion matrices for signal-based (top) and video-based models (bottom).

This demonstrates that the video-based model with 94.05% accuracy outperformed the sensor model with 86.82% accuracy, in terms of stability and classification performance, as mentioned in Table 3. Given identical hardware resources, the CNN-LSTM model took around 40 milliseconds, and the Video Swin-B model took approximately 150 milliseconds on average to carry out a prediction on one test sample.

## 6 Discussions

The proposed study implemented motion intention recognition systems using two kinds of modalities (sEMG and Videos). There was an inherent imbalance between the 8 classes and the "Actual" and "Intention" groups caused due to the varying durations and nature of the tasks. The two proposed models performed well in classifying human activities despite this significant imbalance. This shows the robustness and generalization of these trained models across multiple subjects. The technique of oversampling of intention classes and the application of class weights to the loss function majorly improved the classification ability of the models.

The performance scores from Table 3 clearly state that video-based data could be preferred to sensor signals. However, the signal-based model was around 3.7 times faster in making predictions on test data. This suggests that for time-sensitive applications, e.g., safety hazard monitoring, muscle signals can be a better choice. Unlike video systems, sensors provide consistent input signals even in varying environmental conditions like lighting, visibility, and obstructions. sEMG sensors are attached directly to the skin, making them highly portable compared to a stable camera setup. Hence, in use cases like material transportation, and pavement construction that involve dynamic surrounding factors and long-distance movements of workers, muscle signals can be more efficient. However, with advancements in transformers, video-based methodologies are attaining higher accuracy in action detection with relatively reduced latencies. Videos can capture the full visual context of surroundings, unlike EMG sensors which only collect motion cues from specific target muscles. Therefore, for operations that involve interactions with multiple workers, and complex behaviors, videos could be a more suitable approach.

## 7 Conclusion

In conclusion, the construction industry can be revolutionized by the integration of mutual collaboration between workers and robots. The proposed system conducted experiments with 3 subjects involving sEMG and video data collection followed by data processing. Finally, a hybrid CNN-LSTM model and a fine-tuned Video Swin Transformer model were implemented to handle muscle signal and video sequences respectively. This study includes comprehensive comparisons and discussions on these modalities which conclude that both are competent in their unique ways and there cannot be a universal approach.

Future advancements can involve experiments with a larger subject pool and broader construction tasks to provide more comprehensive validation of the results obtained. This study can be expanded to build a prototype of HRC in construction sites by deploying models on robots with control interfaces. Inference optimization techniques can be integrated to reduce model latency when deployed on edge devices. Fusion model architectures that can exploit the complementary strengths of both modalities can also be considered.